\documentclass[twocolumn,showpacs,preprintnumbers,amsmath,amssymb]{revtex4}

\usepackage{graphicx}
\usepackage{dcolumn}
\usepackage{bm}
\usepackage{color}

\begin{document}


\title{Impact of local stacking on the graphene-impurity interaction: theory and experiments}

\author{F. Hiebel, P. Mallet, J.-Y. Veuillen and L. Magaud}
\affiliation{%
Institut N\'{e}el, CNRS-UJF, BP 166, 38042 Grenoble Cedex 9, France
}

\date{\today}

\begin{abstract}
We investigate the graphene-impurity interaction problem by combining experimental - scanning tunneling microscopy (STM) and spectroscopy (STS) - and theoretical - Anderson impurity model and density functional theory (DFT) calculations - techniques. We use graphene on the SiC$(000\overline{1}) (2\times2)_C$ reconstruction as a model system. The SiC substrate reconstruction is based on silicon adatoms. Graphene mainly interacts with the dangling bonds of these adatoms which act as impurities. Graphene grown on SiC$(000\overline{1}) (2\times2)_C$ shows domains with various orientations relative to the substrate so that very different local graphene/Si adatom stacking configurations can be probed on a given grain. The position and width of the adatom (impurity) state can be analyzed by STM/STS and related to its local environment owing to the high bias electronic transparency of graphene. The experimental results are compared to Anderson's model predictions and complemented by DFT calculations for some specific local environments. We conclude that the adatom resonance shows a smaller width and a larger shift toward the Dirac point for an adatom at the center of a graphene hexagon than for an adatom just on top of a C graphene atom.
\end{abstract}

\maketitle

\section{\label{Introduction}Introduction}
Graphene is a one-atom thick crystal composed of carbon atoms arranged on a honeycomb lattice. From this crystallographic structure with two equivalent carbon atoms - labeled A and B- per unit cell arises a unique low-energy electronic structure. It is characterized by a linear and isotropic dispersion relation around the K (K') corner of the Brillouin zone~\cite{wallace, elecprop}. This structure is referred to as Dirac cones and the low-energy excitations in graphene behave like a 2D gas of massless Dirac fermions~\cite{2DgasDirac, zhangQHE}. Among other properties, high carrier mobilities that have been measured in this system make it promising for applications in electronics~\cite{rev_graphene_elec}.

Except for very specific cases, graphene is in interaction with its environment. Substrate is known to affect the graphene properties~\cite{deHeer_rev_transport,mobility_suspended}
as well as intrinsic or intentionally created defects~\cite{ZhangPRL12, WeeksPRX11, BekyarovaJPhysD12, spinhalfparamagnetism, gas_sensor}. Focusing on the latter, graphene is expected to show unusual behavior when interacting with defects. Going more into details, theory predicts an anomalously small broadening of the adatom level. When considering magnetic impurities, this property makes magnetic moment formation easier~\cite{Uchoa2008}. Moreover, the adatom adsorption site may influence the physics of the Kondo effect~\cite{UchoaPRL11, WehlingPRB10}. The adsorption-site dependency of the impurity level broadening and shifting has been computed~\cite{Uchoa2009, Fano_chinois, fano_wehling, Saha_fano}, providing an impurity adsorption-site signature that can be measured by STM. Yet no experimental observation of this effect has been given so far.

In this study, graphene is obtained by graphitization of SiC~\cite{vanBommel,atlantaconfinement}. We consider the graphene-SiC$(000\overline{1})$ interaction and we show how our results are relevant to the investigation of the graphene-impurity problem. When graphene is grown on SiC$(000\overline{1})$ under ultra high vacuum, two SiC surface reconstructions $(3\times3)$ and $(2\times2)_C$ are found at the interface~\cite{nous_interf}.  In case of the $(3\times3)$ interface, the graphene-substrate interaction is very weak~\cite{nous_interf, nous_G3x3} while it is stronger for the $(2\times2)_C$~\cite{nous_interf, nous_DFT2x2, PSS}.  The $\pi$ states are however always detected around the Fermi level in this system~\cite{nous_interf}.  Here we focus on the SiC$(000\overline{1})(2\times2)_C$ interface.
The atomic structure of the $(2\times2)_C$ reconstruction consists in one Si adatom and one C restatom per unit cell, displaying an empty/filled DB respectively. Our study deals with the hybridization between the Si adatom level - which plays the role of a defect - and the graphene states. Graphene islands being rotated with respect to the substrate, spatial variations of the graphene/adatom stacking are observed. This allows us to experimentally investigate the stacking dependence of the interaction for a given impurity level, which is hardly achievable for deposited adatoms since they usually have a well defined adsoption site~\cite{Guisinger09, Wiesendanger12}. The specificity of STM imaging of graphene on SiC, with high-bias measurements giving access to the interface states and low-bias measurements to the graphene states~\cite{nous_JphysD}, allows us to investigate the impact of the interaction on both the adatom and the graphene states.

In section \ref{param}, we provide information on the sample fabrication, on STM measurements and on the ab initio calculation parameters. 
Section \ref{results} begins with an introductive paragraph, providing more details on
 how graphene on SiC$(000\overline{1})(2\times2)_C$ allows us to study experimentally the graphene-impurity interaction problem. 
We then present our experimental and theoretical results. Low bias STM images display stacking dependent perturbations of the graphene local density of states at low energy, the stronger (weaker) perturbations appearing for top (hollow) graphene/adatom stacking. High bias constant current STM images together with tunneling spectroscopy reveal stacking dependent modifications of the adatom resonance in terms of broadening and shifting. We then show that the Anderson single-impurity model accounts for the general trends we observed experimentally. 
Finally, we complement our study of the graphene/adatom interaction with ab initio calculations for the whole graphene/reconstruction system. Keeping in mind the increased system complexity and the fact that we consider one specific geometry, we concentrate on the general trends of the computed electronic structure. We identify a stacking-dependent impact of graphene on the adatom resonance similar to the one derived from the simple Anderson model. Regarding graphene states, maps of integrated DOS show features similar to the ones observed on low-bias STM images.
\section{\label{param}experiment and Calculation details}
\subsection{Experimental aspects}
The sample preparation and characterization were performed under ultrahigh vacuum. We followed the SiC graphitization procedure presented previously~\cite{nous_interf}. We used n doped $6H$-SiC$(000\overline{1})$ samples priorly cleaned by annealing under a Si flux at $850^{\circ}$C. Annealing steps at increasing temperature up to $1100^{\circ}$C allow us to get a graphene coverage of less than a monolayer. Typical LEED patterns show SiC$(3\times3)$ and SiC$(2\times2)$ spots and a ring-shaped graphitic signal with modulated intensity~\cite{emtsev:155303, nous_interf, revstarke, nous_G3x3}. 

The STM and STS measurements were conducted at room temperature with mechanically cut PtIr tips.
On the surface of the samples, we find domains showing the bare SiC$(000\overline{1})(3\times3)$ reconstruction, graphene monolayer islands on the SiC$(000\overline{1})(3\times3)$ reconstruction ($G/3\times3$) and on the SiC$(000\overline{1})(2\times2)_C$ reconstruction ($G/2\times2$) and also few multilayer islands, similarly to previous studies~\cite{nous_interf, revstarke, nous_G3x3}. In this paper, we will focus on the $G/2\times2$ islands using constant current images and tunneling spectroscopy in the current-imaging-tunneling spectroscopy (CITS) mode. In this latter case, a topographic image is acquired at the stabilization conditions $(V_{Stab}, I_t)$. At each point, the current feedback loop is opened while the sample-surface distance (defined by $(V_{Stab}, I_t)$) is maintained constant, in order to record an I(V) curve. The $dI/dV$ curve is then numerically calculated.
\subsection{ab initio calculation framework}
Calculations are carried out using the VASP code~\cite{vasp}, which is based on the density-functional theory (DFT). We use the generalized gradient approximation~\cite{gga} together with a plane wave basis and ultrasoft pseudopotentials~\cite{pseudopot}. The 4H-SiC substrate is modeled by a slab containing 4 SiC bilayers. The backside of the slab is passivated by hydrogen. An empty space of 8 \AA\ separates the graphene layer from the next SiC slab. We use a plane wave basis cutoff of 211 eV. The ultrasoft pseudopotentials have been extensively tested~\cite{ripples, nous_DFT2x2, nous_defauts}. Integration over the Brillouin zone is carried out within the Monckhorst-Pack scheme, using a $6\times6\times5$ to $18\times18\times1$ grid. All the structures were fully converged, with residual forces smaller than $0.015$ eV/\AA.
The local density of states (LDOS) is calculated by projecting the Kohn-Sham wave functions
onto site-centered spherical harmonic functions within a sphere of radius $r=1.65$ \AA\ centered on the considered site.
\section{\label{results} Results and discussion}
\subsection{From graphene on 6H-SiC$(000\overline{1})(2\times2)_C$ to the impurity problem}
\begin{figure}[!h]
\includegraphics[width=0.5\textwidth]{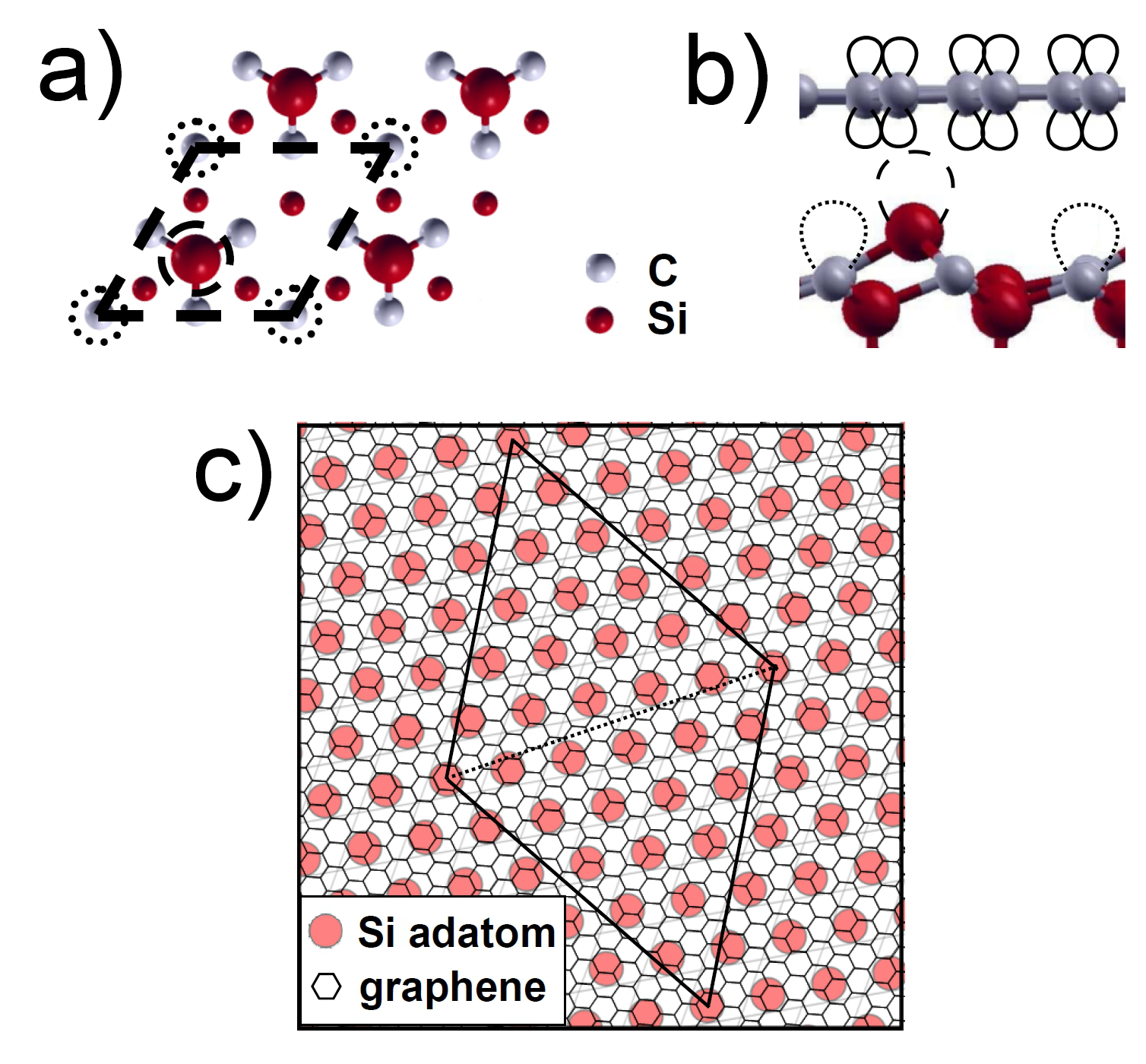}
\caption{\label{fig1} (Color online) (a) Schematic representation of the SiC$(000\overline{1})(2\times2)_C$ surface reconstruction. The unit cell (dashed diamond cell) contains one Si adatom with an empty DB (dashed line) and one C restatom with a filled DB (dotted line). (b) Schematic side view of the system $G/2\times2$ with the $p_z$ graphene orbitals and the empty (filled) DB on the reconstruction adatom (restatom). (c) Schematic top view of the graphene/adatom stacking in case of a rotation angle $\alpha=16\,^{\circ}$. The common pseudocell (black solid line) contains various stacking configurations. }
\end{figure}

The system we study experimentally is composed of a monolayer of graphene on SiC$(000\overline{1})(2\times2)_C$. The atomic structure of the bare substrate reconstruction is now well established and is given in figure \ref{fig1}(a). As it was first proposed from quantitative LEED measurements and STM by Seubert et al.\cite{Seubert}, three out of the 4 dangling bonds of the $(2\times2)$ bulk-truncated cell are saturated by a Si adatom, a priori leaving two half-filled dangling bonds per cell, one on the Si adatom and one on the unsaturated C atom. Later, ab initio calculations~\cite{nous_DFT2x2} along with STM~\cite{PSS, nous_defauts} revealed a charge transfer from the Si adatom to the unsaturated fourth C atom, leaving one empty dangling bond on the Si adatom and one filled dangling bond on the C atom, which we call a restatom. 

In this work, we address the question of the hybridization between graphene $\pi$ states and states from the substrate reconstruction. A schematic side view of the system is given in figure \ref{fig1} (b). We mainly focus on the graphene/adatom interaction since previous ab initio calculations of the band structure of graphene on SiC$(000\overline{1})(2\times2)_C$ have shown that the interaction between graphene and the C restatoms is much weaker than with the Si adatom -- calculations were done either using a GGA~\cite{nous_DFT2x2} or a GGA(D)~\cite{G2x2vdW} exchange-correlation functional, the latter one taking van der Waals interactions empirically into account. This statement is rather intuitive since as depicted in figure \ref{fig1} (b), Si adatoms are much closer to the graphene layer than C restatoms, calculations giving an adatom/restatom height difference of $0.9$ \AA ~\cite{nous_DFT2x2}. 

Another characteristic of the system which is crucial for our study is that on the C face of SiC, graphene grows with a strong rotational disorder under UHV and typical LEED patterns show the graphene signal as a ring with modulated intensity (see EPAPS figure S1 (a)). Figure \ref{fig1} (c) gives a schematic representation of the graphene/adatom stacking for a rotation angle of $\alpha=16\,^{\circ}$. The common pseudocell (given in black) contains various graphene/adatom stacking configurations. The corners of the unit cell display hollow stacking i.e. the adatoms sit under the center of a graphene hexagon. Within the cell, we identify top (shifted top) stacking i.e. the adatoms sit under (close to) a graphene atom or bridge stacking i.e. the adatoms sit under a C-C bond. Unlike in a graphene-adsorbate system, the position of the reconstruction adatom with respect to the graphene lattice is imposed by the interface geometry so that we can investigate the graphene-adatom interaction using room temperature STM measurements. Moreover, adsorbates usually sit on well defined (top, bridge or hollow) energetically favored surface sites~\cite{Guisinger09, Wiesendanger12} which would make the investigation of the stacking-dependence of the interaction hardly possible. In case of $G/2\times2$, spatial variations of the graphene/adatom stacking are imposed by the system and easily identifiable thanks to the transparency of graphene in high bias STM images, as will be discussed in section \ref{results_expe} in connection with figure \ref{fig2}.
Note finally that the adatom states can be considered as independent in a first approximation because they are separated by $6.14$ \AA\ and the graphene/adatom interaction is still sufficiently weak to prevent a coupling through graphene states.
Hence, the system provides the opportunity to investigate how the interaction between graphene and a localized impurity state depends on the local configuration without being limited to stable adsorbate positions.
\subsection{\label{results_expe}Scanning tunneling microscopy and spectroscopy results}
\begin{figure}[!h]
\includegraphics[width=0.5\textwidth]{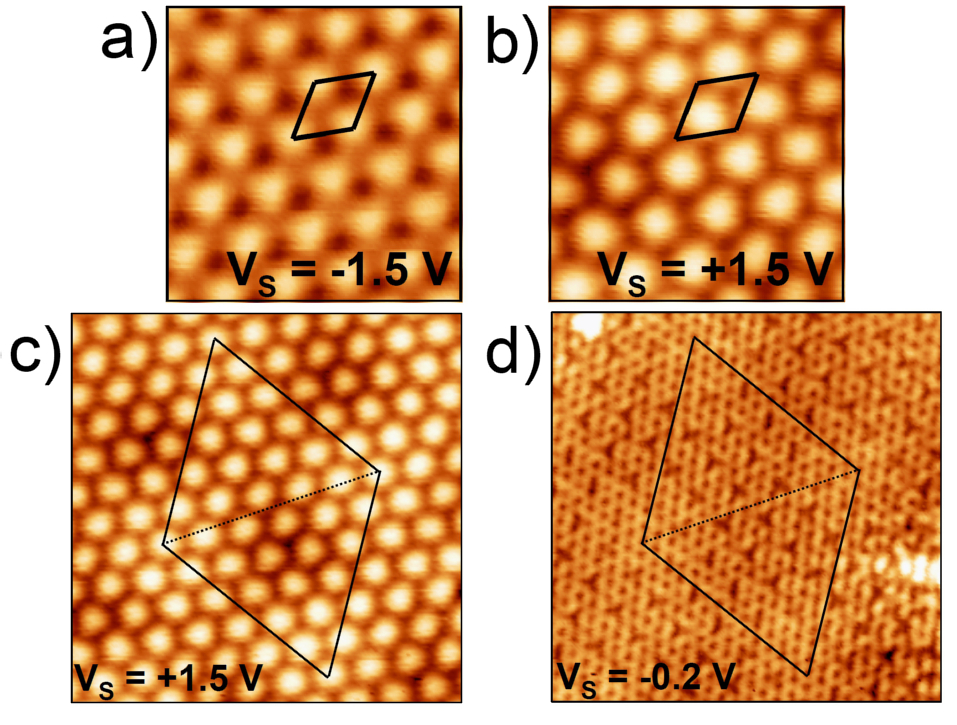}
\caption{\label{fig2} (Color online)(a) and (b) $3\times3$~nm$^2$ Dual-bias STM images of a $G/2\times2$ island. At high bias, images show the substrate reconstruction states (solid line unit cell). (a) ((b)) exhibits the filled (empty) states associated to the restatom (adatom). (c) and (d) $6\times6$~nm$^2$ Dual-bias STM images of the same $G/2\times2$ island as in (a-b). (c) exhibits the adatom empty states while (d) exhibits the graphene overlayer LDOS near $E_F$. A moir\'{e} pattern (solid line cell) is visible on both images (periodicity $P=3.0\pm 0.3$~nm, measured stacking angle $\alpha=16\pm 1\,^{\circ}$). }
\end{figure}

Figure \ref{fig2} (a) and (b) present dual-bias STM images of a $G/2\times2$ island ($V_S=\pm1.5$~V). At such high voltage bias, graphene appears transparent and the images show the reconstruction states, namely with one protrusion per unit cell. At negative (positive) sample bias $V_S$, we probe filled (empty) states. The reference cell (diamond cell) reveals the spatial shift between filled/empty states which is consistent with charge transfer between the reconstruction adatom and restatom predicted by ab initio calculations of the electronic structure of the \itshape bare \upshape SiC$(000\overline{1})(2\times2)_C$ reconstruction~\cite{nous_DFT2x2}. Empty (filled) states are ascribed to states located on the Si adatom (C restatom) from the mentioned ab initio calculations. The electronic structure of the substrate reconstruction is only weakly affected by the graphene layer, indicating a small or moderate graphene-adatom interaction.

Figure \ref{fig2} (c) and (d) are dual-bias STM images of a $G/2\times2$ island at high $V_S>0$ and low $V_S$, respectively showing the reconstruction adatom states and the graphene atomically resolved LDOS near the Fermi level ($E_F$).
On both images, we identify a superstructure of periodicity $P=3.0\pm 0.3$~nm. All the $G/2\times2$ islands observed show such superstructures, however with various periodicities in the nm range. We interpret them as moir\'{e} patterns that arise from the composition of the graphene and the reconstruction lattices stacked with rotation. The geometrical proof of this interpretation, based on the measured relation between the moir\'{e} periodicity and the graphene/substrate rotation angle, is given in the supplementary information and is analogous to the one we applied in \cite{nous_G3x3} to $G/3\times3$ islands. 
Thus, the moir\'{e} pattern (periodicity $P=3.0\pm 0.3$~nm, stacking angle $\alpha=16\pm 1\,^{\circ}$) is visible on both the adatom and the graphene lattice (pseudocell in solid line). On the adatom sites, it appears as a periodic spatial modulation of their apparent height and on the graphene lattice as a periodic repetition of perturbed/unperturbed LDOS zones. The graphene perturbations consist in a LDOS deficit on graphene atoms (``switched off'' atoms) distributed with a periodicity similar to the one of the underlying reconstruction~\cite{nous_interf}. We recall that, from a geometrical point of view, the moir\'{e} pattern is associated to the quasi periodicity of the graphene/reconstruction stacking. Hence, the perturbations observed on both the adatom and graphene lattice being correlated to the moir\'{e} pattern, the graphene substrate/interaction should vary with the local stacking. 

To further study the stacking dependence of the graphene/adatom interaction, we can relate the graphene/adatom stacking and the perturbations using the fact that figure \ref{fig2} (c) and \ref{fig2} (d) are dual bias images. We can thus identify directly the stacking which corresponds to the ``switched off'' graphene atoms. Figure \ref{fig1} (c) in the previous section corresponds to an illustration of the graphene/adatom stacking obtained this way, using figure \ref{fig2} (c-d). On the corners of the pseudocell (in solid line), the adatoms sit under graphene hollow sites. This particular stacking corresponds to small regions ($\approx 1$~nm wide) showing honeycomb contrast typical of graphene on the low bias STM image. Within the two halves of the pseudo cell, adatoms sit under one of the two graphene sites in a top, shifted top or bridge configuration. Those configurations lead to reduced LDOS (``switched off'' atoms) on the concerned graphene sites, in agreement with our previous ab initio calculations \cite{nous_DFT2x2}. Thus, from the point of view of graphene states, the impact of the graphene/adatom interaction is very weak in the hollow stacking configuration while it is stronger in the top, shifted top and bridge configurations. 
In the following, we will concentrate on the effect of this interaction on the adatom states and analyze the moir\'{e} pattern on high bias images.

\begin{figure*}[!t]
\includegraphics[width=\textwidth]{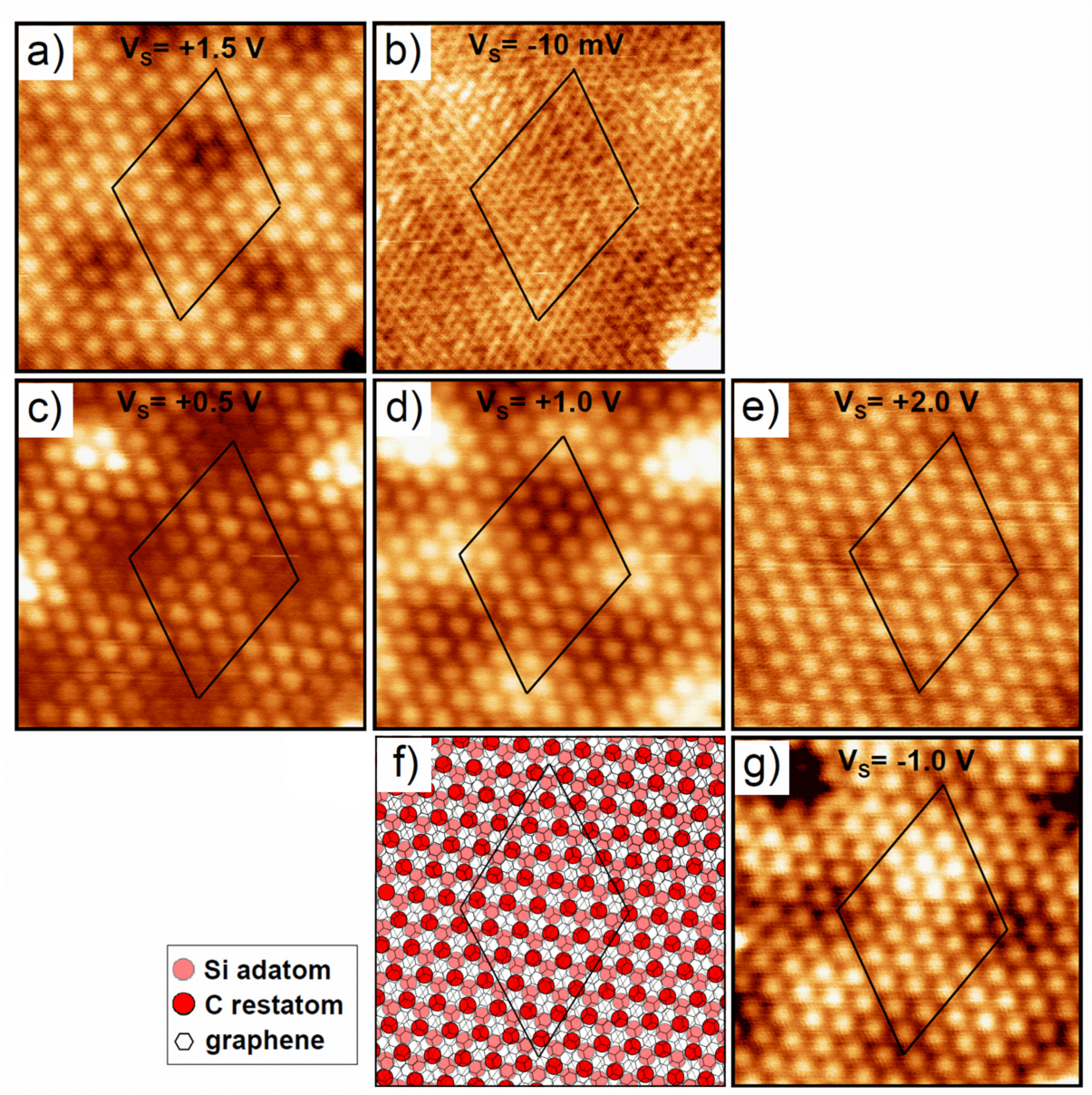}
\caption{\label{fig3} (Color online) Tunneling bias dependance of the moir\'{e} pattern (pseudocell in solid line) contrast, $7\times7$~nm$^2$ constant current mode images. (a) STM image at the reference tunneling bias $V_S=+1.5$~V, (b-e) and (g) being extracted from dual-bias images in order to check the structure positions. (b) Low-bias image showing the unperturbed (corner of the moir\'{e} pseudocell) and pertubed (within the moir\'{e} pseudocell) graphene zones. (c-e) Evolution of the moir\'{e} contrast with $V_S$ on adatom empty states. (d) shows an asymmetry contrast between the top and bottom halves of the moir\'{e} pseudocell. (f) Schematic representation of the graphene/reconstruction stacking deduced from (a), (b) and (g). Honeycomb contrast on (b) corresponds to hollow graphene/adatom and top graphene/restatom stacking. Perturbed graphene regions correspond to top graphene/adatom stacking with top (hollow) graphene/restatom stacking in the top (bottom) half of the moir\'{e} pseudocell.
(g) Moir\'{e} contrast on restatom states (filled states). }
\end{figure*}

STM images contain information on the topography and the electronic density of states of the sample surface. In order to decorrelate those two contributions and investigate the nature of the moir\'{e} pattern that appears when imaging the adatom states, we first focus on the dependence of constant current STM images on the tunneling bias.
Figure \ref{fig3} presents a series of STM images taken at various sample biases on a $G/2\times2$ island. Systematic dual-bias image acquisition at the reference tunneling bias $V_S=+1.5$~V (see figure \ref{fig3} (a)) ensures that images are taken at the same spot of the island and that no tip changes occurred between images. The island shows an incommensurate moir\'{e} pattern (black pseudocell) of pseudoperiodicity $P=3.3\pm0.3$~nm. We measure the rotation angle $\alpha=17\pm1\,^{\circ}$ between the graphene and the $(2\times2)$ lattices using the low bias image ($V_S=10$~mV) in figure \ref{fig3} (b).

The low bias STM image in figure \ref{fig3} (b) ($V_S=10$~mV) furthermore allows us to identify the graphene/adatom stacking and to locate the typical zones of stronger and weaker perturbations of the graphene electronic structure that we presented in connection with figure \ref{fig2} (d). The moir\'{e} pseudocell is represented in black. On the whole series of images the corners of the cell correspond to hollow stacking and to weaker graphene perturbation. Within the cell, we identify top, shifted top and bridge stacking which correspond to stronger graphene perturbation.

The series of high positive bias images showing the adatom states (figures \ref{fig3} (c-e)) exhibits two inversions in the moir\'{e} pattern contrast between $V_S=+0.5$~V and $V_S=+2.0$~V. This behavior indicates that the moir\'{e} pattern is related to a spatial modulation of the adatom LDOS rather than to a topographic modulation. This aspect will be further investigated through scanning tunneling spectroscopy.
As a side remark, one can note the asymmetry between the two halves of the moir\'{e} pseudocell at $V_S=+1.0$~V, the top half appearing darker than the lower one. This differentiation is unlikely to arise from the type of graphene sublattice impacted since they are equivalent in LDOS in ideal graphene. A schematic representation of the graphene/reconstruction stacking given in figure \ref{fig3} (f) helps to find an explanation to this observation. Both halves of the pseudocell show top graphene/adatom stacking at their center, the difference between them lies in the graphene/restatom stacking. The upper (lower) half shows a top (hollow) graphene/restatom stacking. A weak graphene/restatom interaction could thus explain the differentiation between the two halves of the moir\'{e} pseudocell. This assumption is further supported by the differentiation of the restatoms in correlation to the moir\'{e} pattern revealed by the STM image at $V_S=-1.0$~V (filled states) in figure \ref{fig3} (g). 
Finally, this interpretation is consistent with the residual graphene/restatom interaction obtained in recent GGA(D) ab initio calculations for this system~\cite{G2x2vdW}, which authors show as responsible for n-type doping of the graphene layer.

In the following, we focus on the much stronger graphene/adatom interaction and complement data from figure \ref{fig3} with scanning tunneling spectroscopy measurements. The constant current image contrast is generally interpreted as a map of isovalues of the sample LDOS $\rho_S((x, y),E_F)$ for low $V_S$ or of $\rho_S((x, y),E)T(E, V_S, z)$ integrated from $E_F$ to $E_F+eV_S$ for higher tunneling biases, with $T(E, V_S, z)$ the tunnel barrier transmission coefficient. In a spectroscopy measurement, the $dI/dV_S$ is considered as a probe of $\rho_S((x, y),E_F+V_S)$, for slowly varying $T(E, V_S, z)$.

\begin{figure}[!h]
\includegraphics[width=0.5\textwidth]{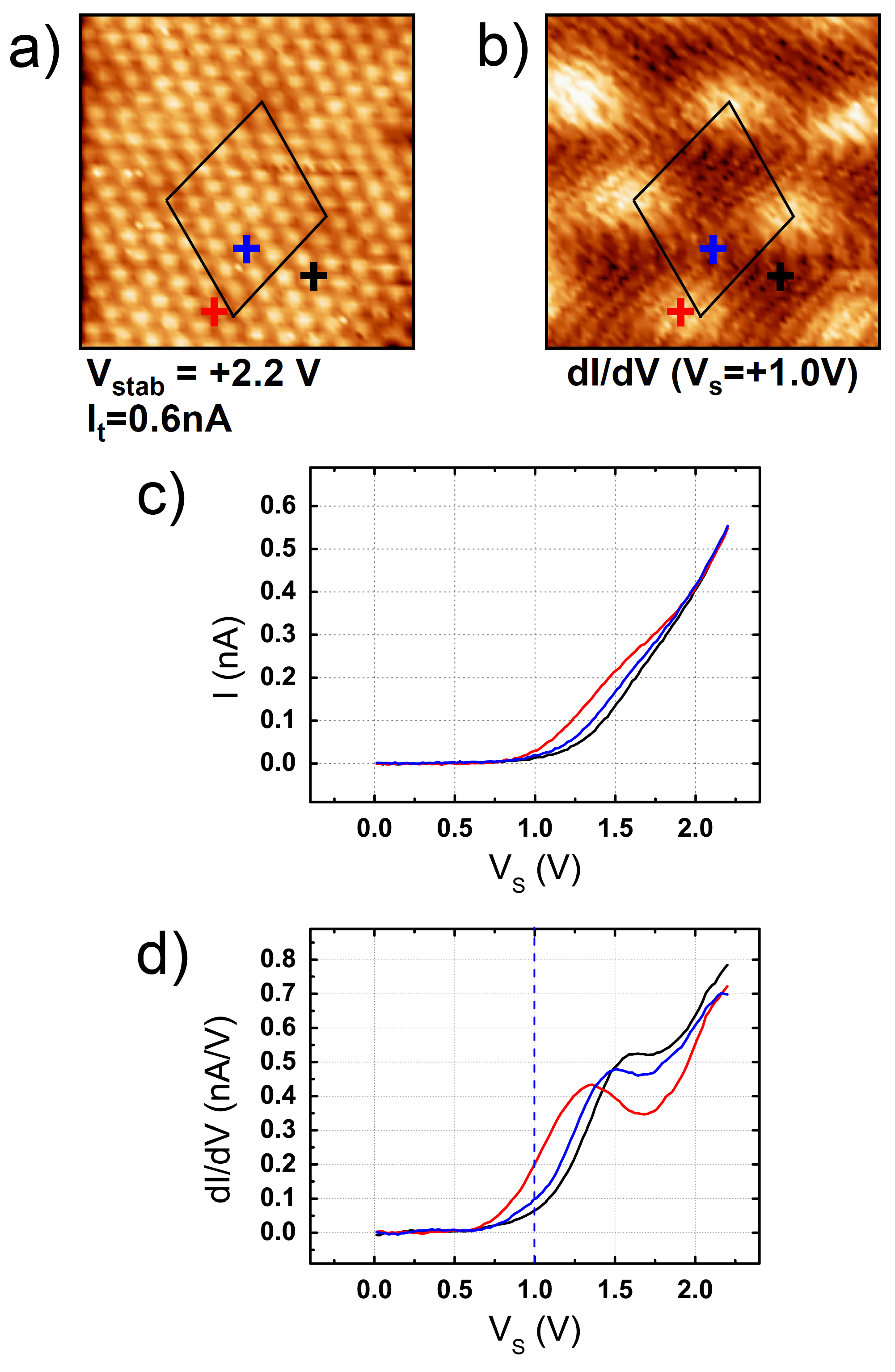}
\caption{\label{fig4} (Color online) CITS data on the same island as in figure \ref{fig3}. (a) $7\times7$~nm$^2$ constant current image at the CITS stabilization parameters  (moir\'{e} pseudocell in black). (b) Conductance image revealing the moir\'{e} pattern. (c) I(V) curves spatially averaged over $2.5$~nm$^2$ ($20$ spectra), on the three different graphene/reconstruction stacking regions (see crosses in (a) and (b): graphene/adatom: hollow (red) graphene/adatom top and graphene/restatom hollow (blue) graphene/restatom top (black).  (d) Corresponding conductance curves. Each curve shows a peaked structure ascribed to the adatom band, at $V_S=+1.35$~V (red), $V_S=+1.52$~V (blue) and $V_S=+1.61$~V (black).}
\end{figure}

Figure \ref{fig4} shows data extracted from a CITS measurement on the same $G/2\times2$ island as in figure \ref{fig3}. The corresponding $7\times7$~nm$^2$ topographic image at stabilization parameters ($V_{stab}=+2.2$~V, $I_t=0.6$~nA) is given in figure \ref{fig4} (a). The $dI/dV$ map at $V_S=+1.0$~V in figure \ref{fig4} (b) clearly reveals a modulation of the surface LDOS in correlation with the moir\'{e} pattern. Figure \ref{fig4} (c) and (d) show $I(V)$ and $dI/dV$ curves acquired within the three different stacking domains (indicated by crosses on figure \ref{fig4} (a-b)) identified using data in figure \ref{fig3}, namely hollow graphene/adatom (red), top graphene/adatom and hollow graphene/restatom (blue) and top graphene/adatom and top graphene/restatom (black). In each spectrum, a clear feature appears at $V_S=+1.35$~V, $V_S=+1.52$~V and $V_S=+1.61$~V respectively~\cite{spectro_norm}, which we ascribe to the adatom resonance. This establishes that the stacking-dependent graphene/adatom interaction shifts the adatom resonance energy. The adatom in top configuration shows the higher resonance energy, which corresponds to configuration leading to the stronger impact on graphene low-energy electronic structure (figure \ref{fig3} (b)). Now going back to the data presented in figure \ref{fig3}, the shifts in the resonance energy do explain the moir\'{e} contrast inversion between $V_S=+1.0$~V and $V_S=+2.0$~V constant current images of figures \ref{fig3} (d) and (e) but does not account for the lower-energy contrast inversion between figures \ref{fig3} (c) and (d). 
This lower-energy contrast inversion could however be interpreted as arising from a stacking dependent resonance broadening along with the stacking dependent shift in energy. The hollow adatom resonance would thus be lower in energy but narrower than the top adatom resonance, allowing us to probe those latter states at energies $\leq0.5$~eV, like in figure \ref{fig3} (c).

The existence of those low-energy states is clear from constant current images but hardly evidenced by CITS data. An explanation to this comes from the fact that each spectrum is acquired at a fixed tip-surface distance which is determined by the stabilization parameters $(V_{Stab}, I_t)$ while during constant current imaging, the tip-sample distance is adjusted in order to keep the tunneling current fixed. As one can infer from the dramatic current drop with $V_S$ in figure \ref{fig4} (c), during the acquisition of the spectra presented in figure \ref{fig4}, the tip-surface distance was relatively high with respect to the one during a constant current image at ($V_S=0.5$~V, $I_t=0.2$~nA) presented in figure \ref{fig3} (c). $Z(V)$ curves on this surface actually indicate that the tip moves toward the surface by $\approx 3$ \AA\ between $V_S=+2.2$~V and $V_S=+0.5$~V (see EPAPS). Thus, low bias states are hardly probed in the CITS measurements of figure \ref{fig4}.

In the following section, we will use a simple theoretical model in order to check if a stacking dependent graphene/adatom interaction can indeed lead to variations in the shifting and broadening of the adatom level.

\subsection{\label{results_anderson}The Anderson single-impurity model}
\begin{figure*}[t]
\includegraphics[width=0.8\textwidth]{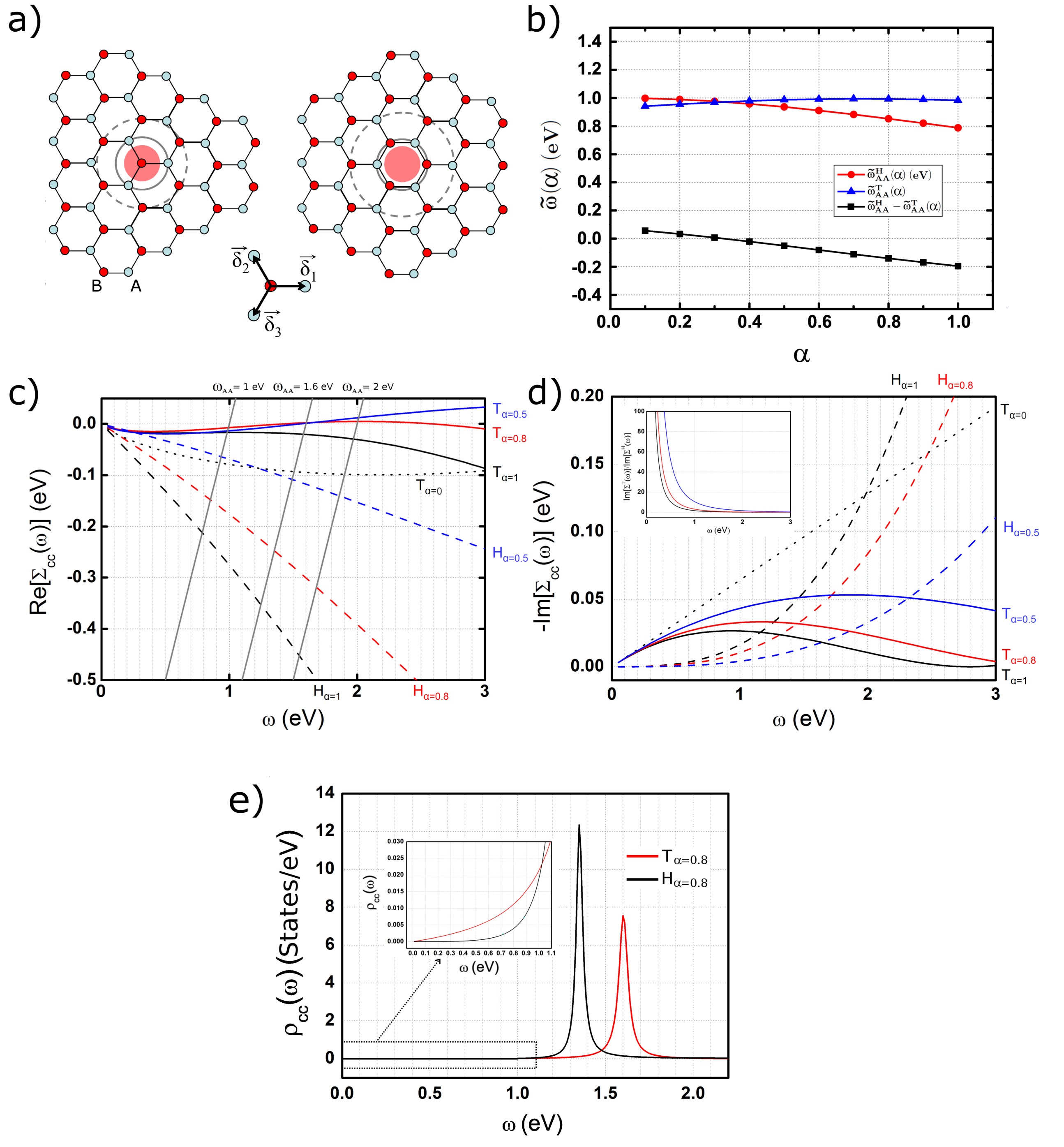}
 \caption{\label{fig5} (Color online) The graphene/adatom interaction in the framework of the Anderson model. (a) The top (left) and hollow (right) configurations. (b) Interacting level energy in both hollow ($\widetilde{\omega}^H_{AA}$) and top ($\widetilde{\omega}^T_{AA}$) configurations and their relative shift $\widetilde{\omega}^H_{AA}-\widetilde{\omega}^T_{AA}$, for $\omega_{AA}= 1$~eV and $V=1$~eV as a function of $\alpha$. (c) Real part of the adatom self-energy in the top ($T_{\alpha}$) and hollow ($H_{\alpha}$) configurations for various $\alpha\geq0.5$ (and $\alpha=0$ in the top configuration) and $V=1$~eV. The gray straight lines represent $\omega-\omega_{AA}$ for $\omega_{AA}=1$~eV, $1.6$~eV and $2$~eV.
(d) Imaginary part of the adatom self-energy in the top ($T_{\alpha}$) and hollow ($H_{\alpha}$) configurations for various $\alpha\geq0.5$ (and $\alpha=0$ in the top configuration) and $V=1$~eV. Insert: $Im[\Sigma^T(\omega)] / Im[\Sigma^H(\omega)]$ for various $\alpha\geq0.5$ (with the same color code as for $Im[\Sigma]$). (e) Normalized DOS associated to the adatom in both top and hollow configurations with $\omega_{AA}=1.6$~eV, $V=1$~eV and $\alpha=0.8$. Insert: Residual DOS at low energy.}
\end{figure*}

In this section, we follow the calculation procedure presented in \cite{Uchoa2008, Uchoa2009} and apply Anderson's description of a localized state in the continuum~\cite{anderson} to our system i.e. graphene on top of an adatom in the top and hollow stacking configuration (see figure \ref{fig5} (a)). In this framework, we model the adatom as an isolated localized state of energy $\omega_{AA}$ and thus neglect the presumably low interaction between adatoms because of the large adatom-adatom distance -$6.14$ \AA- as well as the small impact of the restatoms located further away from the graphene layer.
The spin degree of freedom will also be omitted since we are not dealing with magnetic defects (i.e. we consider the non interacting Anderson model~\cite{WehlingChem.Phys.Lett09}).
Consequently, the total Hamiltonian of the system is given by:
\begin{equation}
{\cal{H}}= {\cal{H}}_g + {\cal{H}}_{AA} +{\cal{H}}_{int} 
\label{}
\end{equation}
${\cal{H}}_g$ is the tight-binding Hamiltonian for isolated graphene in the nearest neighbor approximation:
\begin{equation}
{\cal{H}}_g = - t \sum_{\left\langle i,j\right\rangle}{a^{\dagger}(\mathbf{R_i})b(\mathbf{R_j})+ H. c.}
\label{}
\end{equation}
a, b are the fermionic operators for each A and B graphene sublattice. $\mathbf{R_{i,j}}$	are the atomic positions in the direct space. $t\approx 2.8$~eV is the hopping integral between two nearest neighbors.
In the momentum space, we get:
\begin{equation}
{\cal{H}}_g = - t \sum_{\mathbf{k}}{[\phi(\mathbf{k})a^{\dagger}_{\mathbf{k}}b_{\mathbf{k}}+\phi(\mathbf{k})^*b^{\dagger}_{\mathbf{k}} a_{\mathbf{k}} ]}
\label{}
\end{equation}
with $\phi(\mathbf{k})= \sum_{i=1}^3{e^{i\mathbf{k}\mathbf{\delta}_i}}  $. $\mathbf{\delta}_1= \hat{x}$, $\mathbf{\delta}_2= -\frac{\hat{x}}{2} + \frac{\sqrt{3}}{2}\hat{y}$ and $\mathbf{\delta}_3= -\frac{\hat{x}}{2} - \frac{\sqrt{3}}{2}\hat{y}$ are the vectors of the graphene lattice connecting a $B$-type atom to its three nearest neighbors (see figure \ref{fig5} (a)).
${\cal{H}}_{AA}$ is the Hamiltonian of the unperturbed adatom orbital of energy $\omega_{AA}$ with the fermionic operator $c$ :
\begin{equation}
{\cal{H}}_{AA} = \omega_{AA}c^{\dagger}c
\label{}
\end{equation}
${\cal{H}}_{int}$ is the Hamiltonian that describes the coupling between the adatom s-like orbital -the adatom $s- p_z$ atomic orbital showing axial symmetry- and the graphene states continuum.
${\cal{H}}_{int}$ depends on the graphene/adatom stacking (see table \ref{table_H}).
For the top stacking (${\cal{H}}_{int}={\cal{H}}_{T}$), hopping between the adatom and the graphene sites up to the second neighbors is considered.
For the hollow stacking (${\cal{H}}_{int}={\cal{H}}_{H}$), hopping between the adatom and the six nearest graphene neighbors is considered.
\begin{table}[!h]
\renewcommand{\arraystretch}{1.3}
\centering
\begin{tabular}{|c|c|}
    \hline
${\cal{H}}_{H}$ & $+ \alpha V\sum_{\mathbf{k}}{[\phi(\mathbf{k})b^{\dagger}_{\mathbf{k}}+\phi(\mathbf{k})^*a^{\dagger}_{\mathbf{k}}]c +H. c.}$\\
    \hline
    ${\cal{H}}_{T}$ &  $+ V\sum_{\mathbf{k}}[{b^{\dagger}_{\mathbf{k}}c + \alpha \phi(\mathbf{k})a^{\dagger}_{\mathbf{k}}c + H. c.]}$\\
    \hline
\end{tabular}
\caption{\label{table_H}Graphene/adatom interaction Hamiltonian for the hollow (${\cal{H}}_{H}$) and top (${\cal{H}}_{T}$) stacking configurations.}
\end{table}

 In these expressions, $V$ is the coupling energy between the adatom state and the first-neighbor graphene site in the top configuration.  We introduce the $\alpha \leq1$ parameter in order to take into account the distance dependence of the coupling energy. $\alpha V$  corresponds to the hopping energy between the adatom and the first graphene neighbors in the hollow configuration or the second neighbors in the top configuration (see figure \ref{fig5} (a)).
Taking the second neighbors into account in the top configuration is motivated by the fact that STM images display large adatom orbitals with respect to the graphene lattice and the local graphene/adatom interaction is thus likely to involve more than one graphene atom. Also, this corresponds to the same spatial interaction cut off in the hollow and top configuration (see figure \ref{fig5} (a)). This refinement of the calculation presented in~\cite{Uchoa2008, Uchoa2009} is necessary to treat the graphene adatom interaction on an equal footing for the two stacking configurations we consider here.

We calculate the localized state retarded Green's function $G_{cc}(\tau)=-\left\langle T[c(\tau)c(0)^{\dagger}]\right\rangle$ (with $T$ the time ordering operator) whose expression is:
\begin{equation}
G_{cc}^R(\omega)=[\omega -\omega_{AA}-\Sigma_{cc}(\omega)]^{-1}
\label{}
\end{equation}
with $\Sigma_{cc}(\omega)$ the localized state self-energy arising from the hybridization to the graphene continuum. The hollow ($\Sigma_{cc}^{H} (\omega)$) and top ($\Sigma_{cc}^{T} (\omega)$) configuration self-energies are given in table \ref{table_self}, as functions of matrix components of the bare graphene retarded Green's function $G_{xy, \mathbf{k}}^{0R}(\omega)$ with $x, y=a, b$:

\begin{equation}
G_{xy, \mathbf{k}}^{0R}(\omega) = \frac{\omega\sigma_{xy}^0 - t\sigma_{xy}^1 Re\phi(\mathbf{k}) +t\sigma_{xy}^2Im\phi(\mathbf{k})}{\omega^2 - t^2 |\phi(\mathbf{k})|^2 + i0^+sign(\omega)}
\label{}
\end{equation}
where $\sigma^0$ is the identity matrix and $\sigma^j$ ($j=1, 2$) the real and imaginary off-diagonal Pauli matrices.

\begin{table}[!h]
\renewcommand{\arraystretch}{1.3}
\centering
\begin{tabular}{|c|c|}
   \hline
    $ \Sigma_{cc}^{H} (\omega)$ & $\alpha^2 V^2 \sum_{\mathbf{k}}[\phi(\mathbf{k}) \theta_{a,\mathbf{k}}(\omega) + \phi(\mathbf{k})^* \theta_{b,\mathbf{k}}(\omega)]$~\cite{Uchoa2009} \\
    \ &  with $\theta_{x,\mathbf{k}}(\omega)=\phi(\mathbf{k}) G_{xb, \mathbf{k}}^{0R}(\omega) + \phi(\mathbf{k})^* G_{xa, \mathbf{k}}^{0R}(\omega)$\\
       \hline    
    $\Sigma_{cc}^{T} (\omega)$ & $V^2 \sum_{\mathbf{k}}{[(1+\alpha^2 |\phi(\mathbf{k})|^2 )G_{bb, \mathbf{k}}^{0R}(\omega)}+ 2 \alpha \phi(\mathbf{k}) G_{ba, \mathbf{k}}^{0R}(\omega) ]$ \\
       \hline
\end{tabular}
\caption{\label{table_self} Hollow ($\Sigma_{cc}^{H} (\omega)$) and top ($\Sigma_{cc}^{T} (\omega)$) configuration self-energies as functions of the bare graphene retarded Green's function coefficients $G_{xy, \mathbf{k}}^{0R}(\omega)$.}
\end{table}

We apply the linear-band approximation which is valid for sufficiently low energy and express the energy as $\epsilon_{\pm}(k)=\pm v_F k$ with $v_F\approx 10^6$~m/s around the K(K') point and chose a cut off energy $D\approx7$~eV in order to match the number of states in the first Brillouin zone. We then get the localized state self-energy expressions in table \ref{table_self_lin}:
\newpage
\begin{widetext}
\begin{table}[!t]
\renewcommand{\arraystretch}{1.3}
\centering
\begin{tabular}{|c|c|c|}
   \hline
    $ \Sigma_{cc}^{H} (\omega)$ & $ -\omega \frac{2\Delta \alpha^2}{\pi t^2}[D^2+\omega^2 ln|1-\frac{D^2}{\omega^2}|] - 2 i \Delta \alpha^2 \frac{|\omega|^3}{t^2}\theta(D-|\omega|)$& \cite{Uchoa2008, Uchoa2009}\\
       \hline    
    $\Sigma_{cc}^{T} (\omega)$, $\alpha=0$  & $-\frac{\omega \Delta}{\pi} ln|1-\frac{D^2}{\omega^2}| -i\Delta|\omega|\theta(D-|\omega|)$& \cite{Uchoa2008}\\
       \hline
    $\Sigma_{cc}^{T} (\omega)$ & $ - \frac{\omega \Delta}{\pi t^2}[\alpha^2 D^2 - 4 t \alpha D+2 \alpha t \omega  ln\frac{|D+\omega|}{|D-\omega|}+(\alpha^2 \omega^2 + t^2) ln|1-\frac{D^2}{\omega^2}|] $ & \ \\
    \ & $-i\frac{\Delta}{t^2}[\alpha^2|\omega|^3-2 \alpha t \omega^2 + t^2 |\omega|]\theta(D-|\omega|)$ & \ \\
       \hline
\end{tabular}
\caption{\label{table_self_lin} Hollow ($\Sigma_{cc}^{H} (\omega)$) and top ($\Sigma_{cc}^{T} (\omega)$) configuration localized state self-energies within the linear-band approximation. $\Delta=\pi (\frac{V}{D})^2$ is the dimensionless hybridization parameter.}
\end{table}
\end{widetext}

We are interested in the DOS associated to the adatom, which in this model reads : $\rho _{cc}(\omega)=-\frac{1}{\pi}Im[G_{cc}^R(\omega)]$. The adatom DOS can be expressed in terms of imaginary and real part of the self energy as follows: 
\begin{equation}
\rho _{cc}(\omega)=-\frac{1}{\pi}\frac{Im[\Sigma(\omega)]}{[\omega-\omega_{AA}-Re[\Sigma(\omega)]]^2+[Im[\Sigma(\omega)]]^2}
\label{dos_sigma}
\end{equation}

From equation \ref{dos_sigma}, the shift in energy of the adatom band due to the coupling to the graphene states is related to the real part of the self energy. The renormalized level energy $\widetilde{\omega}_{AA}$ is given by :
\begin{equation}
\widetilde{\omega}_{AA} -\omega_{AA} \simeq Re[\Sigma(\widetilde{\omega}_{AA})]
\label{eq::resigma}
\end{equation}
Figure \ref{fig5} (b) represents the calculated interacting level energy in both hollow ($\widetilde{\omega}^H_{AA}$) and top ($\widetilde{\omega}^T_{AA}$) configurations for $\omega_{AA}= 1$~eV as a function of $\alpha$. For $\alpha \geq 0.3$
, $\widetilde{\omega}^H_{AA}<\widetilde{\omega}^T_{AA}$ which corresponds to the experimental situation. More generally, by solving $Re[\Sigma^H(\widetilde{\omega}_{AA})]=Re[\Sigma^T(\widetilde{\omega}_{AA})]$, we get that for $1$~eV~$\lesssim \widetilde{\omega}_{AA}\lesssim 2$~eV, the switch in the order of the resonances occurs for $0.2\lesssim \alpha \lesssim 0.3$. Thus, our model is consistent with the experimental results when $\alpha$ is sufficiently large. Actually, considering $\alpha\approx1$ is a reasonable choice since STM images (figures \ref{fig2}-\ref{fig3}) show orbitals of axial symmetry on the Si adatoms that are large with respect to the graphene lattice - they cover approximately one graphene hexagon. Consistently, ab initio calculated cross sections of (empty states) integrated DOS (figure 1b in reference \cite{PSS}) show an orbital of axial symmetry with respect to the $SiC[0001]$ axis on the Si adatom, which is large with respect to the graphene cell.
An estimation of $\alpha$ can be drawn from the following considerations. If the distance from the adatom to top first neighbours $d_{top}$ is $2.6$~\AA~$< d_{top} <$~$3.1$~\AA~\cite{nous_DFT2x2,G2x2vdW}, the distance to the hollow first neighbours is $3.0$~\AA~$< d_{hollow} <$~$3.4$~\AA. Which corresponds to a distance increase of $10$ to $15$\%. The overlap integral between  \itshape s \upshape and \itshape p \upshape orbitals evolving like $1/d^2$~\cite{Har89}, we get $0.75<\alpha< 0.85$.\\

In order to gain more insight into how the positions of the resonances depend on the model parameters $\alpha$, $\omega_{AA}$ and $V$, figure \ref{fig5} (c) shows $Re[\Sigma(\omega)]$ for the top and hollow configurations, for various $\alpha\geq0.5$ and $V=1$~eV. By graphically solving equation \ref{eq::resigma} for $1$~eV$\leq\omega_{AA}\leq2$~eV, we find that increasing $\alpha$ or $\omega_{AA}$ increases the resonances energy separation and increasing $\omega_{AA}$ shifts both resonances to higher energies. Finally, $V$ is a scale parameter appearing as $V^2$ in $Re[\Sigma(\omega)]$ (and $Im[\Sigma(\omega)]$) so that an increase in $V$ would dilate the vertical scale of the graph in figure \ref{fig5} (c) by $V^2$.

We now focus on the residual DOS on the adatom away from the resonance at low energy. From equation \ref{dos_sigma}, in this energy range, the DOS behavior is determined by $Im[\Sigma(\omega)]$, which is represented in figure \ref{fig5} (d) for the top and hollow configurations, for various $\alpha\geq0.5$ and $V=1$~eV.
Note that $Im[\Sigma]$ is usually interpreted as the broadening of the impurity level due to the coupling to the continuum. In our case, the situation is more complex than in the case of a continuum of constant DOS with energy as the exotic DOS of graphene introduces an energy dependence of $Im[\Sigma(\omega)]$. Thus, referring to $Im[\Sigma(\omega)]$ as the broadening of the adatom resonance is an abuse of terminology and we will therefore use quotation marks.

Previous studies compared the $\alpha=0$ top configuration case (dotted line) to the $\alpha=1$ hollow configuration case (black dashed line) and extracted an $\omega$ dependence of $Im[\Sigma]$ in the former case and an $\omega^3/t^2$ dependence in the latter case~\cite{Uchoa2009}. The anomalously small ``broadening'' of the adatom band at low energy in the hollow configuration is explained by the appearance of destructive interferences between the different hopping paths connecting three graphene sites of the same sublattice to the adatom site (see figure \ref{fig5} (a)). The adatom in the $\alpha=0$ top configuration impacting only one graphene site, such interferences do not take place and the band ``broadening'' is thus larger~\cite{Uchoa2009, Fano_chinois, fano_wehling, Saha_fano}.
Introducing hopping from the adatom to the second graphene neighbors in the top configuration ($\alpha \neq 0$) leads to the same hopping path symmetry as in the hollow configuration for the three second neighbors. Consequently, the same interference phenomenon happens and contributions up to $\omega^3/t^2$ appear in the expression of $Im[\Sigma_{cc}^{T, \alpha\neq0} (\omega)]$. 
As graphically shown by figure \ref{fig5} (d), the significance of the higher order terms depends on the strength of the coupling to higher order neighbors ($\alpha V$). Increasing alpha decreases (increases) the ``broadening'' of the top (hollow) resonance. The inset graph shows that the ratio $Im[\Sigma^T(\omega)] / Im[\Sigma^H(\omega)]$ diverges at low energy for all considered values of $\alpha$. Hence, we expect that our model accounts for the larger residual DOS for the top adatom than for the hollow adatom at sufficiently low energy.

Figure \ref{fig5} (e) finally shows the density of states associated to the adatom for both configurations, with $\omega_{AA}=1.6$~eV and $\alpha=0.8$. These values have been obtained by solving equation \ref{eq::resigma} for the hollow and top configurations with $\widetilde{\omega}^H_{AA}=1.35$~eV and $\widetilde{\omega}^T_{AA}=1.6$~eV in order to fit the experiment (see figure \ref{fig4}(d)).
The insert displays the calculated DOS at low energy. In this energy range, the DOS is very low with respect to the vicinity of the resonance, which is consistent with the typical evolution of the tip-surface distance as a function of tunneling bias reported in figure S2. When considering the DOS ratio, we find that the DOS on the top adatom clearly dominates the one on the hollow adatom (at $\omega=0.6$~eV, $\rho _{cc}^T/\rho _{cc}^H \approx 7$)~\cite{note_imSigma}.
Thus, three different energy domains emerge, successively where the DOS of the top adatom dominates ($\omega<1.0$~eV in this case), then the DOS from the hollow adatom gains the upper hand ($1.0$~eV$<\omega<1.5$~eV) and again the DOS from the top adatom dominates ($\omega>1.5$~eV). This result is thus consistent with the experimental measurements presented in figure \ref{fig3} showing two moir\'{e} contrast inversions.
In the end, with reasonable values for $\alpha$, $V$ and $\omega_{AA}$, the simple model we used accounts for the variations in level energy and broadening for the two different graphene/adatom stacking configurations expected from the experiments.

To finish, note that by comparing the calculated LDOS to the experimental conductance spectra, we neglect the energy-dependence of the tunnel barrier transmission coefficient $T(E, V_S, z)$ whose effect is to favor tunneling to higher energy empty states i.e. close to the resonances. Also, we assume a weak coupling to the tip states. Moreover, we do not take into account the Fano-resonance effect that can arise from interference between tunneling paths connecting the tip and graphene and the tip and the localized state~\cite{Fano1960, fano_wehling, Saha_fano}. In fact, we expect no such effect here because graphene on SiC is electronically transparent on high bias STM images~\cite{mallet, Strociointerf, nous_JphysD}. Indeed, the STM constant current images of figure \ref{fig3} do not show the graphene atomic contrast for tunneling biases down to $V_S=+0.5$~eV while the adatom states are clearly identified. In other words, at high tunneling bias, direct tunneling from the tip states to the adatom states is highly favored with respect to tunneling from the tip states to the graphene states even though the adatoms lay below the graphene plane.

\subsection{\label{results_ab_initio}The graphene-adatom interaction from ab initio calculations for graphene on SiC$(000\overline{1})(2\times2)_C$}
\begin{figure}[!h]
\includegraphics[width=0.45\textwidth]{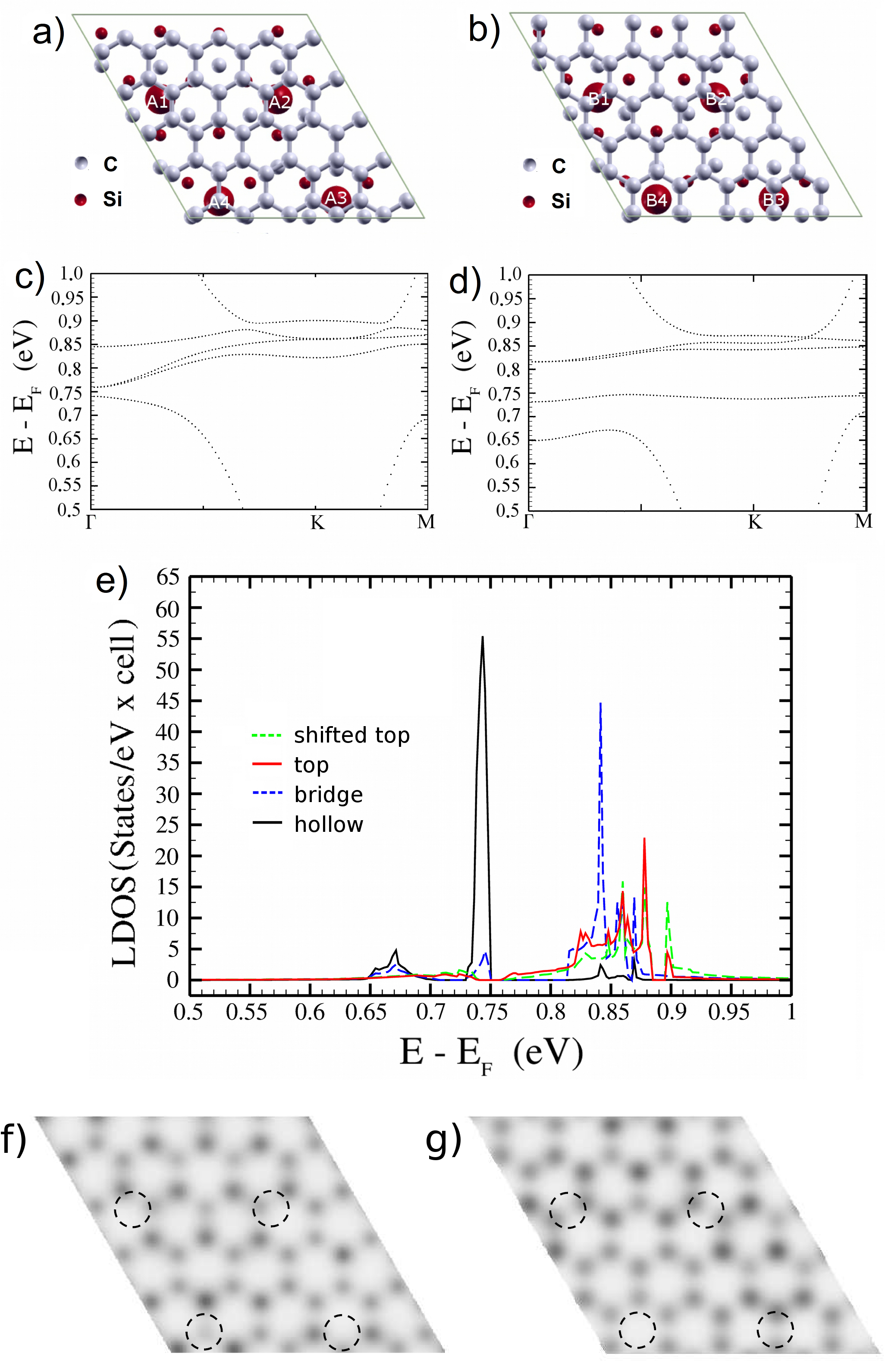}
\caption{\label{fig6} (Color online) Ab initio electronic structure calculations of graphene on SiC$(000\overline{1})(2\times2)$. (a) and (b) Top view of the two considered supercells, displaying top (A4), slightly shifted top (A1-3), bridge (B1-3) and hollow (B4) graphene/adatom stacking configurations. Only the last SiC bilayer, the adatoms and the graphene layer are represented. (c) and (d) respective band structure calculations in the energy range spanning the adatom low-dispersive bands. (e) Calculated LDOS for the A3-4 and B3-4 adatoms. (f), (g) Maps of $|\Psi|^2$ integrated over the energy range $[+0.5$~eV$,+1.0$~eV$]$ taken just above the graphene atoms, for the configurations given in (a) and (b). Contrast is reversed with respect to Fig. \ref{fig2} (d) and \ref{fig3} (b): higher density is darker. The adatom positions are indicated by dashed circles.}
\end{figure}

In this section, we study the graphene-adatom interaction using ab initio calculations that take into account the complete graphene-SiC$(000\overline{1})(2\times2)_C$ system. 
A detailed description of the calculation was published in~\cite{nous_DFT2x2}. Here we focus on the energy range that corresponds to the graphene-adatom interaction.
Due to technical constraints on the calculation supercell size, we chose the quasi-commensurate $5\times5$ graphene on top of a $4\times4$ SiC cell. In order to examine various graphene/adatom stacking, we consider two supercells which correspond to translations of the graphene lattice with respect to the substrate, as shown in figure \ref{fig6} (a) and (b).
With respect to the graphene lattice, the A4 adatom sits in a top configuration, the A1-3 adatoms in a sligthly shifted top configuration, the B4 adatom in a hollow configuration and the B1-3 adatoms in a bridge configuration. 

The supercells we consider correspond to two specific geometries, each one giving rise to a small moir\'{e} pattern. It is different from the geometry probed by STM which gives rise to a large moir\'{e} pattern that contains all the graphene-adatom stacking configurations listed above within one pseudocell. We thus use our calculations to study the graphene-adatom interaction rather than as a model of the system probed experimentally. Moreover, we have to keep in mind that our GGA calculations of the graphene-SiC$(000\overline{1})(2\times2)_C$ may have some imperfections arising from the possible underestimation of the graphene-reconstruction interaction~\cite{G2x2vdW}, which however are not critical for the purpose of this study. 
Consequences of this are the following. Firstly, we cannot discuss the graphene-restatom interaction since our calculations do not give account for this small coupling~\cite{nous_DFT2x2}. The LDA and GGA(D) calculations from the literature show a band anticrossing of $\approx50$~meV~\cite{G2x2vdW} and we have shown experimentally that the graphene-restatom interaction indeed exists in the discussion in connection with figure \ref{fig3}. However, we chose to concentrate on the much stronger graphene-adatom interaction, which we can discuss with our GGA calculations. 
Secondly, we obtain a neutral graphene layer~\cite{nous_DFT2x2} while angle resolved photoemission spectroscopy~\cite{emtsev:155303} and calculations cited above~\cite{G2x2vdW} indicate electron doping of the graphene sheet so that $E_F\approx E_D+0.2$~eV and $E_F\approx E_D+0.35$~eV respectively. Thus, our calculations may not reproduce the position of $E_F$ in this system.
Finally, we obtain a flatter graphene layer ($0.07$~\AA~\cite{nous_DFT2x2}) than calculations~\cite{G2x2vdW} from the literature which show a small corrugation of $0.16$~\AA\ using LDA and $0.27$~\AA\ using GGA(D). We will discuss the consequences of this later in this section. 

The calculated band structure for each configuration is given in figure \ref{fig6} (c) and (d). In both cases, it exhibits a clear band anticrossing between the 4 weakly dispersive adatom bands and the graphene $\pi^*$ band, evidencing interaction between graphene and adatom states.  Note that far from the adatom states, near the Dirac point, our calculations in the same system showed that the linearity of the $\pi$ bands is recovered~\cite{nous_DFT2x2}, which is also the case for GGA(D) and LDA calculations~\cite{G2x2vdW}.

Figure \ref{fig6} (e) gives the calculated adatom LDOS for each type of stacking.
Features displayed by figure \ref{fig6} (e) are complex but the simple model introduced in the previous section will help analyze them. By comparing the two extreme situations, top stacking (red) and hollow stacking (black), we find a lower position in energy of the spectral weight for the hollow than for the top configuration. Regarding the broadening of the adatom band, we find a larger extension in energy in case of the top configuration. Hence, we find the same trends in the impact of the graphene/adatom interaction on the adatom level as with the Anderson single-impurity model.

Considering intermediate configurations which we did not treat with the previous model, the slightly shifted top position leads to a LDOS very similar to the one for the exact top configuration, in energy and band broadening. Such behavior proves that the graphene/adatom interaction is not strictly local. The bridge configuration leads to an average band energy position similar to the the one for the top configuration but with a slightly narrower distribution. 

Going more into details, the fine structure of the ab initio calculated LDOS may arise from various factors that were not taken into account in the previous model. Indeed, in this system we consider an adatom array on a semi-infinite medium, while in the previous section the impurity was isolated. The weak direct adatom/adatom interaction can slightly affect the results. Moreover, each adatom interacts with graphene which is in interaction with the other adatoms. In the end this can result in the multiplication of the peaks in the adatom LDOS by supercell (periodicity) effects.
Since the multiplication of the peaks is governed by the calculation geometry, which is different from the one in the experimentally probed system, we cannot reasonably extract any general information from the fine structure of the calculated LDOS spectra. Moreover, STM measurements were conducted at room temperature, which implies a $~100$~meV broadening in energy of the LDOS features. For instance, the LDOS crossing at $0.69$~eV in \ref{fig6} (e) is unlikely to appear as a contrast inversion in our STM measurements.  

Finally, figures \ref{fig6} (f-g) show maps of $|\Psi|^2$ integrated over the energy range $[+0.5$~eV$,+1.0$~eV$]$, taken just above the graphene atoms, for the configurations given in (a) and (b). 
We chose the energy range where figures \ref{fig6} (c-d) show a clear hybridization between the graphene and adatom states, in order to reveal the impact of the graphene-adatom interaction on the graphene electronic structure in the direct space and compare it to STM data. Note however that the integration energy range in the ab initio calculated maps is different than the one on low bias STM images. Firstly, it is difficult to compare the very low integrated DOS calculated for lower energy with respect to the Dirac point to STM data. Secondly, we recall that in the experimental system, the position of $E_F$ is most likely shifted toward higher energy with respect to the GGA calculation.

Figures \ref{fig6} (f-g) show LDOS reduction on the graphene atom atop of the adatom (figure \ref{fig6} (f)) and equivalent LDOS on the 6 neighboring graphene atoms in the hollow configuration (figure \ref{fig6} (g)).
We thus find the same kind of features we identified on low bias STM images (figure \ref{fig2} (d) and \ref{fig3} (b)). 
Going more into details, on the low bias STM images, the dark contrast not only corresponds to top, but also to side top and to bridge sites of the Si adatoms.
The shape and apparent depth of the ``switched off'' atoms depends on the local stacking. Only for nearly exact hollow site configuration does the perturbation disappear.
First, this is consistent with the fact that the graphene-adatom interaction has a large spatial extension and thus we consider $\alpha\approx 1$ in the Anderson model. It is also consistent with the fact that adatoms in top, side top and bridge configurations are similarly impacted, as indicated by our ab initio study of the adatom LDOS (figure \ref{fig6} (e)). Moreover, we recall that in the Anderson model, it is only when the very specific hollow configuration is reached that the interference between the hopping paths (at low energy) arises, and that the graphene-adatom states hybridization is subsequently strongly reduced.

To finish, we pointed out that our calculations possibly underestimate the graphene-reconstruction interaction and consequently underestimates the graphene corrugation. Our results thus show that the extinction of the LDOS on the graphene atom in top configuration is already present for a flat graphene layer. 
For a corrugated surface, the top graphene atom is closer to the Si adatom by less than $0.5$ \AA\ according to our low bias measurements and ab initio calculations~\cite{G2x2vdW}. This translates in a larger value of $V$ in the Anderson model i.e. an even larger graphene-adatom states hybridization. From the distance dependence of tight binding parameters, $V$ should however increase by less than $25$\%. We recall that STM constant current images represent both the topography and the electronic structure of the surface. In the end, we expect an enhanced contrast on low-bias STM images from the topographic lowering of the top graphene atoms.
Regarding the Si adatom resonances, a $25$\% larger $V$ in the top configuration reduces the hollow-top resonance energy difference - but not enough to invert their positions (see figure \ref{fig5} (c))- and further increases the top resonance ``broadening'' with respect to the hollow resonance (see figure \ref{fig5} (d)). Thus, taking the corrugation into account will not strongly modify the conclusions of this study.
\section{\label{conclusion} Conclusion}
We have been able to investigate the graphene-impurity interaction as a function of the graphene-impurity local stacking. A direct experimental comparative study has been possible thanks to the specificities of graphene on SiC$(000\overline{1})(2\times2)_C$. The substrate reconstruction is composed of one Si adatom and one C restatom per unit cell, and is mainly interacting with graphene through the adatoms. It has a large cell parameter of $6.14$ \AA, allowing us to consider the adatoms as independent in a first approximation. The graphene/adatom interaction is appreciable but not too strong so that the general electronic structure of graphene and the reconstruction are maintained. The rotation between the graphene and the reconstruction lattices imply spatial variations of the local graphene/adatom stacking, allowing us to compare different configurations on the same STM image. Finally, the electronic transparency of graphene on SiC on high bias STM images makes it possible to clearly identify the local stacking and to study both the impact of the interaction on the graphene LDOS - low bias tunneling regime - and on the adatom level - high bias tunneling regime.

Low bias STM images show that the graphene LDOS near the Fermi level is clearly affected in the top-like graphene/adatom stacking regions, graphene atoms atop of an adatom showing strongly reduced LDOS, while perturbations are hardly noticeable in hollow graphene/adatom stacking regions.
Regarding the adatom resonance, a combination of constant current STM images at various high tunneling biases and tunneling spectroscopy measurements reveals a lower resonance energy and smaller broadening in the hollow configuration than in the top configuration.

The single-impurity Anderson model gives consistent results with the experimental observations for the adatom resonances in the top and hollow configurations when the same interaction cut off for both configurations is considered. 
Within this model, the reduced impurity level broadening in the hollow configuration arises from interferences between hopping paths connecting the adatom site and the neighboring graphene atoms within each graphene sublattice.

Finally, we theoretically investigate the graphene-adatom interaction using ab initio calculations that take the whole graphene/reconstruction structure into account. Considering the LDOS on adatoms, from the more complex calculated electronic structure, we extract the same trends in the stacking-dependent impact of the interaction on the adatom level as with the simple Anderson model and STM.

To conclude, our results confirm the importance of local stacking in the graphene-impurity problem. They show that it is possible to discriminate between two local stacking configurations by comparing the impurity level broadening and shifting. The local stacking can also be identified by considering the perturbations in the graphene LDOS.

\begin{acknowledgments}
We thank Didier Mayou and Simone Fratini for valuable discussions.
This work is supported by a computer grant at 
the ACI CIMENT (phynum project); the French ANR
(GRAPHSIC project ANR-07-BLAN-0161, NANOSIM-GRAPHENE project ANR-09-NANO-016-03); a CIBLE, and a RTRA DISPOGRAPH project. FH held a doctoral support from R\'{e}gion Rh\^{o}ne-Alpes. Some figures
are plotted using Xcrysden.
\end{acknowledgments}


\begin{thebibliography}{45}
\bibitem[{not()}]{note_imSigma}
\bibinfo{note}{The difference in ``broadening'' of the levels is small in the
  vicinity of the resonances since in this energy range, for $\alpha=0.8$ and
  $V=1$~eV, $Im[\Sigma^T(\omega)]\approx Im[\Sigma^H(\omega)]$ (see figure
  \ref{fig5} (d)).}
\end{thebibliography}

\end{document}